\documentclass[11pt]{amsart}
\usepackage{amssymb}
\usepackage{graphicx}

\theoremstyle{definition}

\theoremstyle{remark}

\numberwithin{equation}{section}
\begin{document}
\title[Generalized Camassa-Holm Equations]{Derivation of Generalized Camassa-Holm Equations  from Boussinesq-type Equations}


\author[H. A. Erbay]{H. A. Erbay}

\address{Department of Natural and Mathematical Sciences, Faculty of Engineering, Ozyegin University, \\  Cekmekoy 34794, Istanbul, Turkey}
\email{husnuata.erbay@ozyegin.edu.tr}

\author[S. Erbay]{S. Erbay}

\address{Department of Natural and Mathematical Sciences, Faculty of Engineering, Ozyegin University, \\  Cekmekoy 34794, Istanbul, Turkey}
\email{saadet.erbay@ozyegin.edu.tr}

\author[A. Erkip]{A. Erkip}

\address{Faculty of Engineering and Natural Sciences, Sabanci University, \\ Tuzla 34956,  Istanbul,    Turkey}
\email{albert@sabanciuniv.edu}


\date{}

\thispagestyle{empty}

\begin{abstract}
In this paper we derive generalized forms of the Camassa-Holm (CH) equation from  a Boussinesq-type equation  using a two-parameter asymptotic expansion based on two small parameters characterizing nonlinear and dispersive effects and strictly following the arguments in the asymptotic derivation of the classical CH equation. The resulting equations generalize the CH equation   in two different ways. The first generalization replaces the quadratic nonlinearity of the CH equation with a general power-type nonlinearity while the second one replaces the dispersive terms of the CH equation with fractional-type dispersive terms. In the absence of both higher-order nonlinearities and fractional-type dispersive effects, the generalized equations derived reduce to the classical CH equation that describes unidirectional propagation of shallow water waves.  The generalized equations obtained are compared to similar equations available in the literature, and  this leads to the observation that the present equations have not appeared in the literature.
\end{abstract}

\maketitle

\section{Introduction}\label{sec:intro}
 In the present paper, starting from a Bousinesq-type equation and strictly following the arguments in the asymptotic derivation of the celebrated Camassa Holm (CH) equation, we  derive a generalized  CH  equation, with both a general power-type nonlinearity and a fractional-type dispersion,  for small-but-finite amplitude long waves.  The generalized  CH equation derived includes, as a special case, the classical CH  equation
 \begin{equation}
      v_{\tau}+\kappa_{1} v_{\zeta}+3 vv_{\zeta}-v_{\zeta\zeta\tau}=\kappa_{2}(2 v_{\zeta}v_{\zeta\zeta}+vv_{\zeta\zeta\zeta}) \label{ch}
 \end{equation}
 widely recognized in the literature on shallow water waves \cite{camassa, johnson1, ionescu, constantin1, johnson2, lannes}. The CH equation (\ref{ch})  derived as a model for unidirectional propagation of small amplitude shallow water waves   is an infinite-dimensional Hamiltonian system that is completely integrable \cite{constantin3, constantin4}. An interesting property of the CH equation is that for a wide class of initial data  the spatial derivative of the solution becomes unbounded in finite time while the solution remains uniformly bounded and this property  is used to relate the CH equation  to wave breaking in the hydrodynamical interpretation \cite{escher}. Moreover, the CH equation admits non-smooth peakon-type (solitary or periodic) traveling wave solutions  and it was shown that these solutions are orbitally stable \cite{strauss, lenells}.   There are also studies that derive the CH equation as an appropriate model equation for nonlinear dispersive elastic waves (see  \cite{erbay1, erbay2} and the references therein).

 In recent years, various generalizations of the CH equation have appeared in the literature. This growing literature focuses mainly  on modified forms of the CH equation. The typical exercise in those studies is to replace the quadratic nonlinear term $vv_{\zeta}$ in the CH equation with the cubic one $v^{2}v_{\zeta}$ or with a general power-law nonlinearity $v^{p}v_{\zeta}$ without changing the other terms  (see for instance \cite{yin1,yin2,ouyang}). This approach is not fully satisfactory  and it deserves some discussion. Minimally, it seems that whether or not the proposed models to be asymptotically correct is not sufficiently clear. At that point, no one at all could miss the fact that the CH equation is shaped asymptotically by the balance among   nonlinear steepening, linear dispersion and nonlinear dispersion. Of these three effects, nonlinear steepening and linear dispersion are captured by the third and fourth terms, respectively, on the left-hand side of (\ref{ch}) while nonlinear dispersion is captured by the terms on right-hand side of (\ref{ch}). In a certain asymptotic regime, it is natural to expect that the changes in the nonlinear term $vv_{\zeta}$ of (\ref{ch}) (that is, the changes in nonlinear steepening) lead to modifications in the terms on the right-hand side of (\ref{ch}), which characterize nonlinear dispersive effects. Our main impetus in the present study is the need for an asymptotically correct  model (on formal asymptotic grounds) of the CH equation with a general power-type nonlinearity. To this end, starting from a Boussinesq-type equation,  using a double asymptotic expansion in two small parameters and following the approaches in \cite{johnson1,constantin1,johnson2,lannes,erbay1} we derive formally a generalized CH equation under the assumption of a general power-type nonlinearity, which seems not to be in the literature. At this point we should point out that, based on  the integrability, different forms of the modified  CH equation with cubic nonlinearity have appeared in the literature   \cite{fuchssteiner,olver,novikov}. In those studies the starting point is to obtain integrable generalizations of the CH equation using the existence of a bi-Hamiltonian structure \cite{fuchssteiner,olver} or the existence of infinite hierarchies of higher symmetries \cite{novikov}.  However, contrary to the previous studies, the present study has been originally motivated by the desire of the authors to derive a modified CH equation under the assumption of a general power-type nonlinearity by strictly following the arguments in the asymptotic derivation of the classical CH equation. At the end, we find that, even in the case of cubic nonlinearity, the modified CH equation derived here differs from those presented in the literature and a different form of the nonlinear dispersive terms  affects the integrability of it. The second goal of the present study is to derive CH-type equations driven by fractional dispersion. In a recent study \cite{erbay1}, starting from a Boussinesq-type equation, a fractional-type CH equation with quadratic nonlinearity was derived  for small-but-finite amplitude long waves. Our aim here is to extend the prior analysis to a fractional-type CH equation with power-type nonlinearity.

 In this study,  using a two-parameter asymptotic expansion and following  the procedures described in the literature, we show that the unidirectional propagation of small-but-finite amplitude long wave solutions of the generalized fractional improved Boussinesq equation is governed by  the generalized fractional CH equation. In the absence of fractional dispersion, a reduced version of this full equation gives the generalized  CH equation, including the modified CH equation with cubic nonlinearity as a special case.  Furthermore, in the case of the quadratic nonlinearity, the generalized  CH equation reduces to  (\ref{ch}). As a by-product of the present derivation we also derive  generalized forms of both the Korteweg-de Vries (KdV)  equation \cite{korteweg} and the Benjamin-Bona-Mahony (BBM) equation \cite{bbm}. The important point to note here is that the asymptotic derivation of the CH-type equations requires a double asymptotic expansion in two small parameters characterizing nonlinear and  dispersive effects.  However, the asymptotic derivations of the KdV and BBM equations do not necessarily have this property; assuming  that the two parameters are equal, they can be derived using a single-parameter asymptotic expansion as well.

The paper is organized as follows.   In Section \ref{sec:deriv}, introducing slow variables and using a double asymptotic expansion, the generalized fractional   CH equation with both a general power nonlinearity and fractional derivatives  is derived from a Boussinesq-type equation.  In Section \ref{sec:cases},   special cases of the generalized fractional CH equation are discussed and  the unidirectional wave equations derived are presented in the original coordinates.

\section{Derivation of The Generalized Fractional Camassa-Holm Equation}\label{sec:deriv}
In this section we present a formal derivation of the generalized fractional CH equation using a double asymptotic expansion. To this end we study the asymptotic behavior of unidirectional, small-but-finite amplitude, long wave solutions of the  fractional improved Boussinesq equation  with power-type nonlinearities:
\begin{equation}
    u_{tt}-u_{xx}+(-D_{x}^{2})^{\nu}u_{tt}=(u^{p+1})_{xx},  ~~~~p\geq 1,  \label{frac}
\end{equation}
where $p$ is an integer, $\nu$ may not be an integer and the operator $(-D_{x}^{2})^{\nu}$ is defined in terms of the Fourier transform operator ${\mathcal F}$ and its inverse ${\mathcal F}^{-1}$ by $(-D_{x}^{2})^{\nu}q={\mathcal F}^{-1}(|\xi|^{2\nu}{\mathcal F}q)$. In  \cite{duruk1}, the local well-posedness of solutions to the Cauchy problem defined for (\ref{frac}) imposes the restriction $\nu \geq 1$.  For a connection of (\ref{frac}) with nonlinear dispersive elastic waves, we refer the reader to  \cite{erbay1}.   Notice that, when $\nu =1$, (\ref{frac}) reduces to the improved Boussinesq equation
\begin{equation}
    u_{tt}-u_{xx}-u_{xxtt}=(u^{p+1})_{xx}.  \label{imbq}
\end{equation}
Before starting our analysis,  we need to remark upon the following fact. Let us consider  the transformation $q(x)=Q(X)$ with
$X=\delta x$ where $\delta$ is a positive constant. In  \cite{erbay1}, using the Fourier transform and and its inverse, it has been shown that the operator $(-D_{x}^{2})^{\nu}$ scales as $\delta^{2\nu}$ so that $(-D_{x}^{2})^{\nu}q(x)=\delta^{2\nu}(-D_{X}^{2})^{\nu}Q(X)$.

Henceforth, we shall consider only right-going, small-but-finite amplitude, long wave solutions of (\ref{frac}). Assume that $\epsilon>0$ and $\delta>0$ are two small independent parameters, not necessarily of the same order of magnitude. By performing the scaling transformation
 \begin{equation}
    u(x,t)= \epsilon U(Y,S), ~~~~Y=\delta (x-t), ~~~~ S=\delta t \label{approx}
 \end{equation}
to  (\ref{frac}), we obtain
 \begin{equation}
     U_{SS}- 2U_{YS}+\delta^{2\nu}(-D_{Y}^{2})^{\nu}   (U_{SS}-2U_{SY}+U_{YY})= \epsilon^{p}  (U^{p+1})_{YY} \label{per-bous}
 \end{equation}
for $U(Y,S)$. It is obvious from (\ref{per-bous}) that the parameters $\epsilon$ and $\delta$ can be regarded as measures of nonlinear and dispersive effects, respectively. That is, $\epsilon$ can be taken a typical (small) amplitude of waves whereas $\delta$ is taken a typical (small) wavenumber. There is a considerable literature on the asymptotic derivation of the CH equation \cite{johnson1,constantin1,johnson2,lannes,erbay1} and we apply similar ideas to obtain the generalized fractional CH equation. We consider a  double asymptotic expansion of the solution of (\ref{per-bous}) in the form
\begin{equation}
    U(Y,S;\epsilon, \delta)= U_{0}(Y,S)+\epsilon^{p} U_{1}(Y,S)+\delta^{2\nu} U_{2}(Y,S)+\epsilon^{p} \delta^{2\nu} U_{3}(Y,S)
    +{\mathcal O}(\epsilon^{2p},\delta^{4\nu}) \label{sol}
\end{equation}
 as $\epsilon \rightarrow 0$, $\delta \rightarrow 0$. Furthermore we require that the unknowns $U_{n}$ $(n=0,1,2, ...)$ and their derivatives decay to zero as  $|Y| \rightarrow \infty$.  Inserting the asymptotic solution (\ref{sol}) into (\ref{per-bous}) and equating coefficients of the corresponding powers of $\epsilon^{p}$ and $\delta^{2\nu}$,   we get a hierarchy of partial differential equations for the functions  $U_{n}$ $(n=0,1,2, ...)$. The leading-order term $U_{0}$ is governed by the first-order linear partial differential equation
 \begin{equation}
    (D_S-2D_Y) U_{0S} =0. \label{eq-a}
 \end{equation}
Because of our assumption that only the right-going waves will be considered, we have  $U_{0S}=0$ which implies  $U_{0}=U_{0}(Y)$. The next order term $U_{1}$ satisfies
\begin{equation}
    (D_S-2D_Y) U_{1S}  -(U_{0}^{p+1})_{YY}=0 \label{eq-b}
 \end{equation}
at ${\mathcal O}(\epsilon^{p})$.  But, by differentiating this equation with respect to $S$, we notice that $U_{1SS}$ also satisfies (\ref{eq-a}). Again, our assumption on the right-going waves implies that  $U_{1SS}=0$. Using this result we get
\begin{equation}
    U_{1S}=-\frac{1}{2}(U_{0}^{p+1})_Y.  \label{eq-c}
\end{equation}
At order $\delta^{2\nu}$, it is found that, with the use of $U_{0S}=0$, the governing equation for $U_{2}$ is
 \begin{equation}
    (D_S-2D_Y) U_{2S} +(-D_{Y}^{2})^{\nu}U_{0YY}=0. \label{eq-d}
 \end{equation}
Differentiation of (\ref{eq-d}) with respect to $S$ and a similar argument as in (\ref{eq-b}) yield  $U_{2SS}=0$. This implies that
\begin{equation}
    U_{2S}=\frac{1}{2}(-D_{Y}^{2})^{\nu}U_{0Y} \label{eq-e}
 \end{equation}
is the solution of (\ref{eq-d}). At ${\mathcal O}(\epsilon^{p} \delta^{2\nu})$, with the use of $U_{1SS}=0$, we get
 \begin{equation}
    (D_S-2D_Y) U_{3S}  +(-D_{Y}^{2})^{\nu}(U_{1YY}-2 U_{1SY})=(p+1)(U_{0}^{p} U_{2})_{YY}. \label{eq-f}
  \end{equation}
When we differentiate this equation twice with respect to $S$, we see that $U_{3SSS}$ satisfies the same equation as (\ref{eq-a}). Again, our  assumption on the right-going waves implies that $U_{3SSS}=0$. If we differentiate (\ref{eq-f}) with respect to $S$ and substitute (\ref{eq-c}), (\ref{eq-e}) and  $U_{3SSS}=0$ into the resulting equation, we obtain
  \begin{equation}
    U_{3SS}=-\frac{1}{4} (-D_{Y}^{2})^{\nu}(U_{0}^{p+1})_{YY}-\frac{1}{4}(p+1)[U_{0}^{p} (-D_{Y}^{2})^{\nu}   U_{0Y}]_Y. \label{eq-g}
 \end{equation}
Using  this result in  (\ref{eq-f}), $U_{3S}$ is found in the form
 \begin{eqnarray}
    && U_{3S}=-{{p+1}\over 2}(U_{0}^{p} U_{2})_Y+\frac{1}{2}(-D_{Y}^{2})^{\nu} U_{1Y}+\frac{3}{8} (-D_{Y}^{2})^{\nu}(U_{0}^{p+1})_{Y} \label{eq-k}  \\
          && \quad \quad \quad -\frac{p+1}{8} U_{0}^{p}(-D_{Y}^{2})^{\nu} U_{0Y}.   \nonumber 
  \end{eqnarray}
 For our purposes, the higher-order terms in the asymptotic expansion will not be needed in the later analysis.

 Now, differentiating both sides of (\ref{sol}) with respect to $S$ once and using all the results obtained above in the resulting equation we get
 \begin{eqnarray}
 \label{eq-h}  \\
    &&   ~U_S =\epsilon^{p}  U_{1S} +\delta^{2\nu}  U_{2S}+ \epsilon^{p}\delta^{2\nu}  U_{3S}+{\mathcal O}(\epsilon^{2p},\delta^{4\nu}) \nonumber  \\
     &&  \quad \quad  = -\frac{\epsilon^{p}}{2}  \left[ U_{0}^{p+1} + (p+1)\delta^{2\nu} U_{0}^{p} U_{2}\right]_Y
            +\frac{\delta^{2\nu}}{2}(-D_{Y}^{2})^{\nu} \left(U_{0} + \epsilon^{p} U_{1}\right)_{Y} \nonumber \\
        && \quad \quad   +\frac{\epsilon^{p} \delta^{2\nu}(p+1)}{8} \left[3(-D_{Y}^{2})^{\nu}(U_{0}^{p}U_{0Y})- U_{0}^{p} (-D_{Y}^{2})^{\nu}U_{0Y}\right]
                +{\mathcal O}(\epsilon^{2p},\delta^{4\nu}).     \nonumber    
 \end{eqnarray}
 Using $U_{0S}=0$, (\ref{eq-c}) and  (\ref{eq-e}) in (\ref{eq-h}) and   keeping all terms up to ${\mathcal O}(\epsilon^{p},\delta^{2\nu})$, we get
 \begin{eqnarray}
 \label{eq-i} \\
   && U_S+{\epsilon^{p}\over 2} (U^{p+1})_Y -\frac{\delta^{2\nu}}{2}(-D_{Y}^{2})^{\nu} U_{Y}
        -\frac{\epsilon^{p} \delta^{2\nu} (p+1)}{8} \left[3(-D_{Y}^{2})^{\nu}(U^{p}U_{Y}) \right.   \nonumber \\
    && \quad \quad \left. - U^{p} (-D_{Y}^{2})^{\nu} U_{Y}\right]=0  \nonumber
\end{eqnarray} 
 for $U(Y,S;\epsilon, \delta)$. Notice that, for $p=1$ and $\nu=1$, this equation reduces to
\begin{equation}
    U_S+{\epsilon\over 2} (U^{2})_Y +\frac{\delta^{2}}{2}U_{YYY}
        +\frac{\epsilon \delta^{2}}{4} \left[3(UU_{Y})_{YY} - U U_{YYY}\right]=0. \label{eq-i-ch}
\end{equation}
The crucial observation for (\ref{eq-i-ch}) is that it is not the CH equation in its standard form. In Ref. \cite{erbay1}, (\ref{eq-i-ch}) has been converted into the standard form of the CH equation in two stages. First, (\ref{eq-i-ch}) has been rewritten in a moving frame of reference.  Second, to incorporate the term $U_{YYS}$ of the CH equation into (\ref{eq-i-ch}), a classical trick which was proposed to derive the BBM   equation \cite{bbm} from the KdV  equation \cite{korteweg} has been performed. In the remaining part of this section this approach will be extended to (\ref{eq-i}).

We first consider the moving frame
\begin{equation}
    X=aY+bS, ~~~~T=cS \label{eq-j}
\end{equation}
where $a$, $b$ and $c$ are positive constants to be determined later. And we rewrite (\ref{eq-i}) in the new coordinate system with $U(Y,S)=V(aY+bS, cS)=V(X,T)$:
\begin{eqnarray}
  &&  cV_T+b V_X+{{a\epsilon^{p}}\over 2}(V^{p+1})_X -\frac{\delta^{2\nu} a^{2\nu +1}}{2} (-D_{X}^{2})^{\nu}V_{X} \label{eq-k}  \\
  && \quad  \quad -\frac{(p+1)\epsilon^{p} \delta^{2\nu}  a^{2\nu +1}}{8} \left[3(-D_{X}^{2})^{\nu}(V^{p}V_{X})
                        - V^{p}(-D_{X}^{2})^{\nu} V_{X}\right]=0. \nonumber
\end{eqnarray}
Then, inserting the relation
 \begin{equation}
  (-D_{X}^{2})^{\nu}V_{X}=-{c\over b}(-D_{X}^{2})^{\nu}V_{T}
                -{{a\epsilon^{p}}\over {2b}}(-D_{X}^{2})^{\nu} (V^{p+1})_{X}+{\mathcal O}(\delta^{2\nu},\epsilon^{p}\delta^{2\nu}) \label{eq-l}
 \end{equation}
obtained from  (\ref{eq-k}) into again (\ref{eq-k}), we eliminate the term $(-D_{X}^{2})^{\nu}V_{X}$ in favor of $(-D_{X}^{2})^{\nu}V_{T}$.   Thus, (\ref{eq-k}) becomes
\begin{eqnarray}
  \label{eq-m} \\
  && \!\!\!\!\!\!\!\! V_T+{b\over c} V_X+{{a\epsilon^{p}}\over {2c}} (V^{p+1})_X +\frac{\delta^{2\nu} a^{2\nu +1}}{2b}(-D_{X}^{2})^{\nu} V_{T} \nonumber \\
  && \quad  -\frac{(p+1)\epsilon^{p} \delta^{2\nu} a^{2\nu +1}}{8c}\left[(3-\frac{2a}{b})(-D_{X}^{2})^{\nu}(V^{p}V_{X})
                        - V^{p} (-D_{X}^{2})^{\nu}V_{X}\right]=0.  \nonumber
\end{eqnarray}
Inserting the scaling transformation
 \begin{equation}
    v=\epsilon V, ~~~~X=\delta \zeta, ~~~~T=\delta \tau \label{eq-n}
\end{equation}
into (\ref{eq-m}) remove the parameters $\epsilon$ and $\delta$ from (\ref{eq-m}) and gives
\begin{eqnarray}
    && v_\tau+{b\over c} v_\zeta+{a\over {2c}} (v^{p+1})_\zeta +\frac{a^{2\nu +1}}{2b} (-D_{\zeta}^{2})^{\nu}v_{\tau}\label{eq-o} \\
    && \quad \quad -\frac{(p+1)a^{2\nu +1}}{8c}\left[(3-\frac{2a}{b})(-D_{\zeta}^{2})^{\nu}(v^{p}v_{\zeta})
                 - v^{p}(-D_{\zeta}^{2})^{\nu} v_{\zeta}\right]=0.  \nonumber
\end{eqnarray}
We note that, for $p=1$ and $\nu=1$, this equation reduces to
\begin{equation}
    v_\tau+{b\over c} v_\zeta+{a\over {2c}} (v^{2})_\zeta -\frac{a^{3}}{2b} v_{\zeta\zeta\tau}
        +\frac{a^{3}}{4c}\left[(3-\frac{2a}{b})(vv_{\zeta})_{\zeta\zeta} - v v_{\zeta\zeta\zeta}\right]=0.  \label{eq-o-ch}
\end{equation}
In  \cite{erbay1} it has been shown that this equation can be converted to the CH equation (\ref{ch}) with special values of  $\kappa_{1}$ and $\kappa_{2}$ when the parameters $a$, $b$ and $c$ are chosen such that the following three conditions are satisfied. The first condition requires that the ratio of the coefficients of the terms $v_\zeta v_{\zeta\zeta}$ and $v v_{\zeta\zeta\zeta}$  must be 2:1. The second and third conditions say that the coefficients of the terms $vv_{\zeta}$ and $v_{\tau\zeta\zeta}$ are 3 and -1, respectively. These conditions are widely used in the literature \cite{johnson1,constantin1,johnson2} and  we refer the reader to \cite{lannes} for more detailed discussions on these conditions. We now follow a similar procedure to fix the free parameters $a$, $b$ and $c$ appearing in (\ref{eq-o}). Under similar conditions for (\ref{eq-o}) we get
\begin{equation}
    a=\left({2\over \sqrt{5}}\right)^{1/\nu},~~~~b={2\over 5}\left({2\over {\sqrt{5}}}\right)^{1/\nu},~~~~c={1\over 3}\left({2\over {\sqrt{5}}}\right)^{1/\nu}. \label{param-a}
 \end{equation}
Notice that, for these values of $a$, $b$ and $c$, (\ref{eq-m})  and (\ref{eq-o}) reduce to
\begin{eqnarray}
     && V_{T}+\frac{6}{5}V_{X}+{3\over 2}\epsilon^{p} (V^{p+1})_{X}+\delta^{2\nu} (-D_{X}^{2})^{\nu}  V_{T}  \label{eq-r} \\
      && \quad \quad +\frac{3(p+1)\epsilon^{p} \delta^{2\nu}}{10}\left[2(-D_{X}^{2})^{\nu}(V^{p}V_{X}) +V^{p}(-D_{X}^{2})^{\nu}V_{X}\right]=0, \nonumber
\end{eqnarray}
and
\begin{equation}
      v_{\tau}+{6\over 5} v_{\zeta}+{3\over 2} (v^{p+1})_{\zeta}+(-D_{\zeta}^{2})^{\nu}v_{\tau}
        =-{{3(p+1)}\over 10}\left[2(-D_{\zeta}^{2})^{\nu}(v^{p}v_{\zeta})+v^{p}(-D_{\zeta}^{2})^{\nu}v_{\zeta}\right],  \label{eq-s}
 \end{equation}
respectively. We henceforth call (\ref{eq-s})  the generalized fractional CH equation because for $p=1$ it reduces to the fractional CH equation presented in  \cite{erbay1} (see equation (4.10) of  \cite{erbay1}). Furthermore, when both $p=1$ and $\nu=1$, (\ref{eq-s}) reduces to the CH equation (\ref{ch}) with $\kappa_{1}={6/5}$ and $\kappa_{2}={9/5}$.  To the best of our knowledge (\ref{eq-s}) has never appeared in the literature prior to the present work. The main distinction between the generalized  fractional CH equation, (\ref{eq-s}), and previous works on the CH-type equations is that (\ref{eq-s}) includes both power nonlinearities and fractional dispersion.

We close this section by stating (\ref{eq-s}) in the original coordinates $x$ and $t$. We first observe that (\ref{approx}), (\ref{eq-j}), (\ref{eq-n}) and (\ref{param-a}) provide the coordinate transformation between $(\zeta, \tau)$ and $(x,t)$ in the form
\begin{equation}
    \zeta=\left({2\over \sqrt{5}}\right)^{1/\nu}(x-{3\over 5}t), ~~~~  \tau={1\over 3}\left({2\over \sqrt{5}}\right)^{1/\nu}t.  \label{trans}
\end{equation}
Using this coordinate transformation and introducing  $w(x, t)=v(\zeta, \tau)$, we convert (\ref{eq-s}) to
\begin{eqnarray}
    &&  w_{t}+ w_{x}+ {1\over 2}(w^{p+1})_{x}+{3\over 4}(-D_{x}^{2})^{\nu}w_{x}+{5\over 4}(-D_{x}^{2})^{\nu}w_{t} \label{ch-or}  \\
    &&  \quad \quad =-{{p+1}\over 8}[2  (-D_{x}^{2})^{\nu}   (w^{p}w_{x})+w^{p} (-D_{x}^{2})^{\nu}  w_{x}]. \nonumber
 \end{eqnarray}
In the following section we will take a closer look at some special cases of the generalized fractional CH equation.

\section{Some Special Cases}\label{sec:cases}
In this section we state some important particular cases of the generalized fractional CH equation (\ref{eq-s}). We also derive generalized forms of the  KdV \cite{korteweg} and BBM \cite{bbm} equations. From the point of view of the KdV and BBM equations, the main difference between our approach and  similar studies in the literature is that the generalized equations derived here include both power nonlinearities and fractional dispersion.

\subsection{The generalized Camassa-Holm equation}
 When $\nu=1$, (\ref{eq-s}) reduces to the generalized CH equation
 \begin{equation}
      v_{\tau}+{6\over 5} v_{\zeta}+{3\over 2} (v^{p+1})_{\zeta}-v_{\zeta\zeta\tau}
        ={{3(p+1)}\over 10}\left[2(v^{p}v_{\zeta})_{\zeta\zeta}+v^{p}v_{\zeta\zeta\zeta}\right].  \label{eq-s-g}
 \end{equation}
Even though various generalizations of the CH equation have been proposed in the literature, this equation is different from  those studied thus far because of the different form of the terms on the right-hand side of (\ref{eq-s-g}). Using the coordinate transformation (\ref{trans}), we can write (\ref{eq-s-g}) in the original reference frame as follows
\begin{equation}
      w_{t}+ w_{x}+ {1\over 2}(w^{p+1})_{x}-{3\over 4}w_{xxx}-{5\over 4}w_{xxt}={{p+1}\over 8}[2  (w^{p}w_{x})_{xx}+w^{p} w_{xxx}] \label{ch-or-g}
 \end{equation}
with $w(x, t)=v(\zeta, \tau)$.

\subsection{The modified Camassa-Holm equation}
When $p=2$, (\ref{eq-s-g}) reduces to the modified CH equation
 \begin{equation}
      v_{\tau}+{6\over 5} v_{\zeta}+{9\over 2} v^{2}v_{\zeta}-v_{\zeta\zeta\tau}
        ={9\over 10}\left[2(v^{2}v_{\zeta})_{\zeta\zeta}+v^{2}v_{\zeta\zeta\zeta}\right].  \label{eq-s-m}
 \end{equation}
 Although there is a growing literature on modified forms of the CH equation, this equation seems to be new in the literature. For instance, the modified CH equation proposed in  \cite{fuchssteiner,olver}, which is an integrable generalization of the classical CH equation, includes extra terms on the right-hand side.   With the use of the coordinate transformation (\ref{trans}) we can rewrite   (\ref{eq-s-m}) in terms of the original variables as
\begin{equation}
      w_{t}+ w_{x}+ {3\over 2}w^{2}w_{x}-{3\over 4}w_{xxx}-{5\over 4}w_{xxt}={3\over 8}[2  (w^{2}w_{x})_{xx}+w^{2} w_{xxx}] \label{ch-or-g}
 \end{equation}
with $w(x, t)=v(\zeta, \tau)$.

\subsection{The generalized fractional BBM equation}
We now  seek an asymptotic solution of (\ref{per-bous}) in the form
 \begin{equation}
    U(Y,S;\epsilon, \delta)= U_{0}(Y,S)+\epsilon^{p} U_{1}(Y,S)+\delta^{2\nu} U_{2}(Y,S) +{\mathcal O}(\epsilon^{2p},\epsilon^{p}\delta^{2\nu}, \delta^{4\nu}) \label{sol-a}
 \end{equation}
This means that the terms of order $\epsilon^{p} \delta^{2\nu}$ are neglected in all calculations in the previous section. Then,  instead of (\ref{eq-s}), we get  the generalized fractional BBM equation
\begin{equation}
      v_{\tau}+\kappa_{1} v_{\zeta}+{3\over 2} (v^{p+1})_{\zeta}+(-D_{\zeta}^{2})^{\nu}v_{\tau}=0,  \label{eq-s-f}
 \end{equation}
 with $\kappa_{1}=3a^{2\nu}/2$ (where $a$ is an arbitrary positive constant). It is interesting to note that, in the special case  $p=1$, (\ref{eq-s-f}) reduces to one discussed in   \cite{kapitula}.  In the original reference frame $(x,t)$,  (\ref{eq-s-f})  becomes
 \begin{equation}
       w_{t}+ w_{x}+ {1\over 2}(w^{p+1})_{x}+{3\over 4}(-D_{x}^{2})^{\nu}w_{x}+{5\over 4}(-D_{x}^{2})^{\nu}w_{t}=0
      \label{bbm-or}
 \end{equation}
with $w(x, t)=v(\zeta, \tau)$  and $a=(4/5)^{1/{(2\nu)}}$.

\subsection{The generalized fractional  KDV equation}
Similarly, using (\ref{sol-a}) instead of (\ref{per-bous}) and neglecting all the terms of $\epsilon^{p} \delta^{2\nu}$, we reach the following equation
 \begin{equation}
    U_S+{\epsilon^{p}\over 2} (U^{p+1})_Y -\frac{\delta^{2\nu}}{2}(-D_{Y}^{2})^{\nu} U_{Y}=0
         \label{eq-i-f}
 \end{equation}
instead of  (\ref{eq-i}). We call (\ref{eq-i-f}) the generalized fractional KdV equation because for $p=1$ it reduces to the fractional KdV equation that is well analyzed in the literature, see, for instance,    \cite{bona}.  In the original reference frame $(x,t)$, (\ref{eq-i-f})  takes the following form
 \begin{equation}
     w_{t}+ w_{x}+ {1\over 2}(w^{p+1}){x}-{1\over 2}(-D_{x}^{2})^{\nu} w_{x}=0
         \label{kdv-or}
 \end{equation}
with $w(x, t)=v(\zeta, \tau)$.


\begin{thebibliography}{00}

\bibitem{bbm}  T. B. Benjamin, J. L.  Bona, and J. J. Mahony,
                    Model equations for long waves in nonlinear dispersive systems, \textit{Philos. Trans. R. Soc. Lond. Ser. A: Math. Phys. Sci.} \textbf{272} (1972)  47--78.
\bibitem{bona}  J. L. Bona, P. E. Souganidis, and W. A. Strauss,
                    Stability and instability of solitary waves of Korteweg-de Vries type, \textit{Proc. R. Soc. Lond. A} \textbf{411} (1987) 395--412.
\bibitem{camassa} R. Camassa and D. D. Holm,
                 An integrable shallow-water equation with peaked solitons, \textit{Phys. Rev. Lett.} \textbf{71} (1993)  1661--1664.

\bibitem{constantin4} A. Constantin,
                    On the scattering problem for the Camassa-Holm equation, \textit{Proc. R. Soc. Lond. A} \textbf{457} (2001) 953--970.

\bibitem{constantin3} A. Constantin and H. P. McKean, A shallow water equation on the circle, \textit{Commun. Pure Appl. Math.} \textbf{52} (1999)  949--982.

\bibitem{escher} A. Constantin and J. Escher,
                    Wave breaking for nonlinear nonlocal shallow water equations, \textit{Acta Math.} \textbf{181} (1998) 229--243.

\bibitem{constantin1}  A. Constantin and D. Lannes,
                    The hydrodynamical relevance of the Camassa-Holm and Degasperis-Procesi equations,   \textit{Arch. Rational Mech. Anal.} \textbf{192} (2009) 165--186.

\bibitem{strauss} A. Constantin and W. A. Strauss,
                    Stability of peakons, \textit{Comm. Pure Appl. Math.} \textbf{53} (2000) 603--610.

\bibitem{duruk1} N. Duruk, H. A. Erbay, and A. Erkip,
            Global existence and blow-up for a class of nonlocal nonlinear  Cauchy problems arising in elasticity, \textit{Nonlinearity} \textbf{23}  (2010)   107--118.

\bibitem{erbay1} H. A. Erbay, S. Erbay, and A. Erkip, Derivation of the Camassa-Holm equations for elastic waves, \textit{Phys. Lett. A} \textbf{379} (2015) 956--961.

\bibitem{erbay2} H. A. Erbay, S. Erbay, and A. Erkip, The Camassa-Holm equation as the long-wave limit of the improved Boussinesq equation and of a class of  nonlocal wave equations, \textit{Discrete Contin. Dyn. Syst.} (to appear) (arXiv:1601.02154v1[math.AP]).

\bibitem{fuchssteiner} B. Fuchssteiner, Some tricks from the symmetry-toolbox for nonlinear equations: Generaliziations of the Camassa-Holm equation, \textit{Physica D} \textbf{95} (1996) 229-243.

\bibitem{ionescu} D. Ionescu-Kruse,
                     Variational derivation of the Camassa-Holm shallow water equation, \textit{J. Non-linear Math. Phys.} \textbf{14} (2007) 303--312.
\bibitem{johnson1}  R. S. Johnson,
                    Camassa-Holm, Korteweg-de Vries and related models for water waves, \textit{J. Fluid Mech.}  \textbf{455} (2002) 63--82.

\bibitem{johnson2}  R. S. Johnson,
                    A selection of nonlinear problems in water waves, analysed by perturbation-parameter techniques, \textit{Commun. Pure Appl. Anal.} \textbf{11} (2012)  1497--1522.

\bibitem{kapitula} T. Kapitula and A. Stefanov,
                    A Hamiltonian-Krein (instability) index theory for solitary waves to KdV-like eigenvalue problems,
                    \textit{Stud. Appl. Math.} \textbf{132} (2014) 183--211.

 \bibitem{korteweg} D. J. Korteweg and G. de Vries,
                    On the change of form of long waves advancing in a rectangular channel, and on a new type of long stationary waves, \textit{Phil. Mag.} \textbf{39} (1895)  422--443.

\bibitem{lannes}  D. Lannes,
                  \textit{The Water Waves Problem: Mathematical Analysis and Asymptotics},  AMS Mathematical Surveys and Monographs \textbf{188} (American Mathematical Society, Providence, Rhode Island, 2013).

\bibitem{lenells} J. Lenells,
                    A variational approach to the stability of periodic peakons, \textit{J. Nonlinear Math. Phys.} \textbf{11} (2004) 151-–163.

\bibitem{ouyang} Z. Liu and Z. Ouyang, A note on solitary waves for modified forms of Camassa-Holm and Degasperis-Procesi equations, \textit{Phys. Lett. A}  \textbf{366} (2007) 377-381.

\bibitem{novikov} V. Novikov, Generalizations of the Camassa-Holm equation, \textit{J. Phys. A: Math. Theor.} \textbf{42}  (2009) 342002 (14pp).

\bibitem{olver} P. J. Olver and P. Rosenau, Tri-Hamiltonian duality between solitons and solitary-wave solutions having compact support,  \textit{Phys. Rev. E} \textbf{53} (1996) 1900-1906.

\bibitem{yin1} Z. Yin, On the blow-up scenario for the generalized Camassa-Holm equation, \textit{Commun. on Partial Differential Equations} \textbf{29} (2004) 867--877.

\bibitem{yin2} Z. Yin, On the Cauchy problem for the generalized Camassa-Holm equation, \textit{Nonlinear Analysis} \textbf{66} (2007) 460-471.

\end{thebibliography}
\end{document}